\documentclass[11pt, letterpaper]{amsart}
\usepackage{graphicx, amssymb}

\newtheorem{thm}{Theorem}[section]

\newtheorem{lem}[thm]{Lemma}
\newtheorem{prop}[thm]{Proposition}
\theoremstyle{definition}

\theoremstyle{remark}
\newtheorem{rem}[thm]{Remark}

\numberwithin{equation}{section}
\newcommand{\norm}[1]{\left\Vert#1\right\Vert}
\newcommand{\abs}[1]{\left\vert#1\right\vert}
\newcommand{\set}[1]{\left\{#1\right\}}
\newcommand{\Real}{\mathbb R}
\newcommand{\Natural}{\mathbb N}

\newcommand{\such}{\ | \ }
\newcommand{\prob}{\mathbb{P}}
\newcommand{\Exp}{\mathcal E}
\newcommand{\parti}{\mathbb{T}}

\newcommand{\expec}{\mathbb{E}}

\newcommand{\basis}{(\Omega, \, \mathcal{F}, \, \pare{\mathcal{F}_t}_{t \in \Real_+}, \, \prob)}
\newcommand{\filtration}{\pare{\mathcal{F}_t}_{t \in \Real_+}}
\newcommand{\F}{\mathcal{F}}

\newcommand{\cadlag}{c\`adl\`ag}

\newcommand{\ud}{\, \mathrm d}
\newcommand{\inner}[2]{\left \langle #1 , #2 \right \rangle}
\newcommand{\dya}{\mathbb{D}}

\newcommand{\Def}{\mathcal{Y}_{\mathsf{s}}}

\newcommand{\simplex}{\triangle^d}

\newcommand{\pare}[1]{\left(#1\right)}
\newcommand{\bra}[1]{\left[#1\right]}
\newcommand{\dbra}[1]{[\kern-0.15em[ #1 ]\kern-0.15em]}
\newcommand{\dbraco}[1]{[\kern-0.15em[ #1 [\kern-0.15em[}
\newcommand{\dbraoc}[1]{]\kern-0.15em] #1 ]\kern-0.15em]}

\newcommand{\X}{\mathcal{X}_{\mathsf{s}}}

\newcommand{\indic}{\mathbb{I}}

\newcommand{\iin}{i \in \set{1, \ldots, d}}

\newcommand{\oX}{\mathcal{X}}
\newcommand{\tx}{\widetilde{X}}
\newcommand{\td}{\widetilde{Y}}

\newcommand{\naone}{NA$1_{\mathsf{s}}$}
\newcommand{\naonep}{NA$1'_{\mathsf{s}}$}

\newcommand{\num}{num\'eraire}

\newcommand{\normtwo}[1]{\norm{#1}_{\mathbb{L}^2}}

\begin{document}

\title[On the semimartingale property of discounted asset-price processes]{On the semimartingale property of discounted asset-price processes}%
\author{Constantinos Kardaras}%
\address{Constantinos Kardaras, Mathematics and Statistics Department, Boston University, 111 Cummington Street, Boston, MA 02215, USA.}%
\email{kardaras@bu.edu}%
\author{Eckhard Platen}%
\address{Eckhard Platen, School of Finance and Economics \& Department of Mathematical Sciences, University of Technology, Sydney, P.O. Box 123, Broadway, NSW 2007, Australia.}%
\email{eckhard.platen@uts.edu.au}%

\thanks{The authors would like to thank the editor, associate editor, and two referees for their careful reading and valuable suggestions that improved the paper. The first author would also like to thank the warm hospitality of the \emph{School of Finance and Economics} of the \emph{University of Technology, Sydney}, where this work was carried out.}%
%\subjclass{60H05 � 60H30 � 91B28}%

\date{\today}%
\dedicatory{This work is dedicated to the memory of our colleague and dear friend Nicola Bruti Liberati, who died tragically on the 28th of August, 2007.}%
%\dedicatory{This is a draft and incomplete version of the paper. Please do not refer to any of the results until the finalized version becomes available.}%
%\commby{}%
%----------------------------------------------------------------
\begin{abstract}
A financial market model where agents trade using realistic combinations of \emph{buy-and-hold} strategies is considered. Minimal assumptions are made on the discounted asset-price process --- in particular, the semimartingale property is \emph{not} assumed. Via a natural market viability assumption, namely, absence of arbitrages of the first kind, we establish that discounted asset-prices \emph{have} to be semimartingales. In a slightly more specialized case, we extend the previous result in a weakened version of the Fundamental Theorem of Asset Pricing that involves strictly positive supermartingale deflators rather than Equivalent Martingale Measures.
\end{abstract}

\maketitle

% ----------------------------------------------------------------
\setcounter{section}{0}

\section{Introduction}

In the process of obtaining a sufficiently general version of the Fundamental Theorem of Asset Pricing (FTAP), semimartingales proved crucial in modelling discounted asset-price processes. The powerful tool of stochastic integration with respect to general predictable integrands, that semimartingales are exactly tailored for, finally lead to the culmination of the theory in \cite{MR1304434}. The FTAP connects the economical notion of \textsl{No Free Lunch with Vanishing Risk} (NFLVR) with the mathematical concept of existence of an \textsl{Equivalent Martingale Measure} (EMM), i.e., an auxiliary probability, equivalent to the original (in the sense that they have the same impossibility events), that makes the discounted asset-price processes have some kind of martingale property. For the above approach to work one has to utilize stochastic integration using \emph{general} predictable integrands, which translates to allowing for \emph{continuous-time} trading in the market. Even though continuous-time trading is of vast theoretical importance, in practice it is only an ideal approximation; the only feasible way of trading is via \emph{simple}, i.e., combinations of \emph{buy-and-hold}, strategies.

Recently, it has been argued that existence of an EMM is \emph{not} necessary for viability of the market; to this effect, see \cite{MR1774056}, \cite{MR1929597}, \cite{MR2210925}. Even in cases where classical arbitrage opportunities are present in the market, credit constraints will not allow for arbitrages to be scaled to any desired degree. It is rather the existence of a \textsl{strictly positive supermartingale deflator}, a concept weaker than existence of an EMM, that allows for a consistent theory to be developed.

\smallskip

Our purpose in this work is to provide answers to the following questions:

\begin{enumerate}
  \item Why are semimartingales important in modeling discounted asset-price processes?
  \item Is there an analogous result to the FTAP that involves weaker (both economic and mathematical) conditions, and  only assumes the possibility of simple trading?
\end{enumerate}

A partial, but precise, answer to question (1) is already present in \cite{MR1304434}. Roughly speaking, market viability already imposes the semimartingale property on discounted asset-price processes. In this paper, we elaborate on the previous idea, undertaking a different approach, which ultimately leads to an improved result. We also note that in \cite{MR2139030}, \cite{MR2157199} and \cite{LarZit05}, the semimartingale property of discounted asset-price processes is obtained via the finite value of a utility maximization problem; this approach will also be revisited.

All the conditions that have appeared previously in the literature are only \emph{sufficient} to ensure that discounted asset-price processes are semimartingales. Here, we shall also discuss a necessary and sufficient condition in terms of a natural market-viability notion that parallels the FTAP, under minimal initial structural assumptions on the discounted asset-price processes themselves. The weakened version of the FTAP that we shall come up with as an answer to question (2) above is a ``simple, no-short-sales trading'' version of Theorem 4.12 from \cite{MR2335830}.

\smallskip

The structure of the paper is as follows. In Section \ref{sec: UPBR, supdefl, semimarts, FTAP}, we introduce the market model, simple trading under no-short-sales constraints. Then, we discuss the market viability condition of \textsl{absence of arbitrages of the first kind} (a weakening of condition NFLVR), as well as the concept of \textsl{strictly positive supermartingale deflators}. After this, our main result, Theorem \ref{thm: FTAP_jump}, is formulated and proved, which establishes both the importance of semimartingales in financial modelling, as well as the weak version of the FTAP. Section \ref{sec: beyond main result} deals with remarks on, and ramifications of, Theorem \ref{thm: FTAP_jump}.
We note that, though hidden in the background, the proofs of our results depend heavily on the notion of the \emph{\num \ portfolio} (also called \emph{growth-optimal}, \emph{log-optimal} or \emph{benchmark portfolio}), as it appears in a series of works: \cite{KELLY}, \cite{LONG}, \cite{MR1849424}, \cite{MR1970286}, \cite{MR1929597}, \cite{MR2194899}, \cite{MR2335830}, \cite{MR2284490}, to mention a few.

\section{The Semimartingale Property of Discounted Asset-Price Process and a Version of the Fundamental Theorem of Asset Pricing} \label{sec: UPBR, supdefl, semimarts, FTAP}

\subsection{The financial market model and trading via simple, no-short-sales strategies}

The random movement of $d \in \Natural$ risky assets in the market is modelled via \cadlag,
%\ (right-continuous with left-hand limits)
nonnegative stochastic processes $S^i$, where $i \in \set{ 1, \ldots, d}$.
%(In most cases, $S^i$, $i \in \set{ 1, \ldots, d}$, denotes the discounted cum-dividend share price process of some company with limited liability, which ensures its nonnegativity.)
As is usual in the field of Mathematical Finance, we assume that all wealth processes are discounted by another special asset which is considered a ``baseline''. The above process $S = (S^i)_{i = 1, \ldots, d}$ is defined on a filtered probability space $\basis$, where $\filtration$ is a filtration satisfying $\F_t \subseteq \F$ for all $t \in \Real_+$, as well as the usual assumptions of right-continuity and saturation by all $\prob$-null sets of $\F$.

Observe that there is no \emph{a priori} assumption on $S$ being a semimartingale. This property will come as a consequence of a natural market viability assumption.

\smallskip

In the market described above, economic agents can trade in order to reallocate their wealth. Consider a \textsl{simple} predictable process $\theta := \sum_{j=1}^n \vartheta_{j} \indic_{\dbraoc{\tau_{j-1}, \tau_j}}$. Here, $\tau_0 = 0$, and for all $j \in \set{ 1, \ldots, n}$ (where $n$ ranges in $\Natural$), $\tau_j$ is a \emph{finite} stopping time and $\vartheta_{j} = (\vartheta^i_j)_{i=1, \ldots, d}$ is $\F_{\tau_{j-1}}$-measurable. Each $\tau_{j-1}$, $j \in \set{ 1, \ldots, n}$, is an instance when some given economic agent may trade in the market; then, $\vartheta^i_j$ is the number of units from the $i$th risky asset that the agent will hold in the trading interval $]\tau_{j-1}, \tau_{j}]$. This form of trading is called \textsl{simple}, as it comprises of a finite number of \textsl{buy-and-hold} strategies, in contrast to \textsl{continuous} trading where one is able to change the position in the assets in a continuous fashion. This last form of trading is only of theoretical value, since it cannot be implemented in reality, even if one ignores market frictions.
%, as we do here to keep the exposition simple.
Starting from initial capital $x \in \Real_+$ and following the strategy described by the simple predictable process $\theta := \sum_{j=1}^n \vartheta_j \indic_{\dbraoc{\tau_{j-1}, \tau_j}}$, the agent's discounted wealth process is given by
\begin{equation} \label{eq: Xhat}
X^{x, \theta} \ = x + \int_0^\cdot \inner{\theta_t}{\ud S_t} \ := \ x + \sum_{j = 1}^n \inner{\vartheta_j}{ S_{\tau_j \wedge \cdot} - S_{\tau_{j-1}  \wedge \cdot}}.
\end{equation}
(We use ``$\inner{\cdot}{\cdot}$'' throughout to denote the  usual Euclidean inner product on $\Real^d$.)

The wealth process $X^{x, \theta}$ of \eqref{eq: Xhat} is \cadlag \ and adapted, but could in principle become negative. In real markets, some economic agents, for instance pension funds, face several institution-based constraints when trading. The most important constraint is prevention of having negative positions in the assets; we plainly call this  \emph{no-short-sales} constraints. In order to ensure that no short sales are allowed in the risky assets, which also include the baseline asset used for discounting, we define $\X (x)$ to be the set of all wealth processes $X^{x, \theta}$ given by \eqref{eq: Xhat}, where $\theta = \sum_{j=1}^n \vartheta_j \indic_{\dbraoc{\tau_{j-1}, \tau_j}}$ is simple and predictable and such that $\vartheta^i_j \geq 0$ and $\inner{ \vartheta_j}{S_{\tau_{j-1}}} \leq X^{x, \theta}_{\tau_{j-1}}$ hold for all  $\iin$ and $j \in \set{ 1, \ldots, n}$. (The subscript ``$\mathsf{s}$'' in $\X(x)$ is a mnemonic for ``simple''; the same is true for all subsequent definitions where this subscript appears.) Note that the previous no-short-sales restrictions, coupled with the nonnegativity of $S^i$, $\iin$, imply the stronger $\theta^i \geq 0$ for all $i = 1, \ldots, d$ and $\inner{ \theta}{ S_-} \leq X^{x, \theta}_-$. (The subscript ``${}_-$'' is used to denote the left-continuous version of a \cadlag \ process.) It is clear that $\X(x)$ is a convex set for all $x \in \Real_+$. Observe also that $\X (x) = x \X (1)$ for all $x \in \Real_+ \setminus \set{0}$. Finally, define $\X := \bigcup_{x \in \Real_+} \X(x)$.

\subsection{Market viability} \label{subsec: UPBR}

We now aim at defining the essential ``no-free-lunch'' concept to be used in our discussion. For $T \in \Real_+$, an $\F_T$-measurable random variable $\xi$ will be called an \textsl{arbitrage of the first kind on $[0, T]$} if $\prob[\xi \geq 0] = 1$, $\prob[\xi > 0] > 0$, and \emph{for all $x > 0$ there exists $X \in \X(x)$, which may depend on $x$, such that $\prob[X_T \geq \xi] = 1$}. If, in a market where only simple, no-short-sales trading is allowed, there are \emph{no} arbitrages of the first kind on \emph{any} interval $[0, T]$, $T \in \Real_+$ we shall say that condition \naone \ holds. It is straightforward to check that condition \naone \ is weaker than condition NFLVR (appropriately stated for simple, no-short-sales trading). The next result describes an equivalent reformulation of condition \naone \ in terms of boundedness in probability of the set of outcomes of wealth processes, which is essentially condition ``\textsl{No Unbounded Profit with Bounded Risk}'' of \cite{MR2335830} for all finite time-horizons in our setting of simple, no-short-sales trading.

\begin{prop} \label{prop: NAone iff NUPBR}
Condition \emph{\naone} holds if and only if, for all $T \in \Real_+$, the set $\set{X_T \such X \in \X(1)}$ is bounded in probability, i.e., $\downarrow \lim_{\ell \to \infty}  \sup_{X \in \X (1)} \, \prob [X_T > \ell] = 0$ holds for all $T \in \Real_+$.
\end{prop}

\begin{proof}
Using the fact that $\X(x) = x \X(1)$ for all $x > 0$, it is straightforward to check that if an arbitrage of the first kind exists on $[0, T]$ for some $T \in \Real_+$ then $\set{X_T \such X \in \X(1)}$ is not bounded in probability. Conversely, assume the existence of $T \in \Real_+$ such that $\set{X_T \such X \in \X(1)}$ is not bounded in probability. As $\set{X_T \such X \in \X(1)}$ is further convex, Lemma 2.3 of \cite{MR1768009} implies the existence of $\Omega_u \in \F_T$ with $\prob[\Omega_u] > 0$ such that, for all $n \in \Natural$, there exists $\tx^n \in \X(1)$ with $\prob[\{\tx^n_T \leq  n \} \cap \Omega_u] \leq \prob[\Omega_u] / 2^{n+1}$. For all $n \in \Natural$, let $A^n = \indic_{\{\tx^n_T > n\}} \cap \Omega_u \in \F_T$. Then, set $A := \bigcap_{n \in \Natural} A^n \in \F_T$ and $\xi :=  \indic_A$. It is clear that $\xi$ is $\F_T$-measurable and that $\prob[\xi \geq 0] = 1$. Furthermore, since $A \subseteq \Omega_u$ and
\[
\prob \bra{\Omega_u \setminus A} = \prob \bra{\bigcup_{n \in \Natural} \pare{\Omega_u \setminus A^n}} \leq \sum_{n \in \Natural} \prob \bra{ \Omega_u \setminus A^n} = \sum_{n \in \Natural} \prob \bra{ \{ \tx^n_T \leq  n \} \cap \Omega_u} \leq \sum_{n \in \Natural} \frac{\prob[\Omega_u]}{2^{n+1}} = \frac{\prob[\Omega_u]}{2},
\]
we obtain $\prob[A] > 0$, i.e., $\prob[\xi > 0] > 0$. For all $n \in \Natural$ set $X^n := (1 / n) \tx^n$, and observe that $X^n \in \X(1 / n)$ and $\xi = \indic_{A} \leq \indic_{A^n} \leq X^n_T$ hold for all $n \in \Natural$. It follows that $\xi$ is and arbitrage of the first kind on $[0, T]$, which finishes the proof.
\end{proof}

\begin{rem} \label{rem: NAone for stop times}
The constant wealth process $X \equiv 1$ belongs to $\X(1)$. Then, Proposition \ref{prop: NAone iff NUPBR} implies that  condition \naone \ is also equivalent to the requirement that the set $\set{X_T \such X \in \X(1)}$ is bounded in probability for all finite stopping times $T$.
\end{rem}

\subsection{Strictly positive supermartingale deflators} \label{subsec: supermart_defl}

Define the set $\Def$ of \textsl{strictly positive supermartingale deflators for simple, no-short-sales trading}  to consist of all \cadlag \ processes $Y$ such that $\prob[Y_0 = 1, \text{ and } Y_t > 0 \ \, \forall t \in \Real_+] = 1$, and $Y X$ is a supermartingale for all $X \in \X$. Note that existence of a strictly positive supermartingale deflator is a condition closely related, but strictly weaker, to existence of equivalent (super)martingale probability measures.

\subsection{The main result} \label{subsec: main result}

Condition \naone, existence of strictly positive supermartingale deflators and the semimartingale property of $S$ are immensely tied to each other, as will be revealed below.

\smallskip

Define the (first) \textsl{bankruptcy time} of $X \in \X$ to be $\zeta^X := \inf \{ t \in \Real_+ \such X_{t-} = 0 \text{ or } X_t = 0\}$. We shall say that $X \in \X$ \textsl{cannot revive from bankruptcy} if $X_t = 0$ holds for all $t \geq \zeta^X$ on $\{ \zeta^X < \infty\}$. As $S^i \in \X$ for $i \in \set{ 1, \ldots, d}$, the previous definitions apply in particular to each $S^i$, $i \in \set{ 1, \ldots, d}$.

Before stating our main Theorem \ref{thm: FTAP_jump}, recall that $S^i$, $i \in \set{1, \ldots, d}$, is an \textsl{exponential semimartingale} if there exists a semimartingale $R^i$ with $R^i_0 = 0$, such that $S^i = S^i_0 \Exp(R^i)$  where ``$\Exp$'' denotes the \textsl{stochastic exponential} operator.

\begin{thm} \label{thm: FTAP_jump}
Let $S = (S^i)_{i=1, \ldots, d}$ be an adapted, \cadlag \ stochastic process such that $S^i$ is nonnegative for all $i \in \set{1, \ldots, d}$. Consider the following four statements:
\begin{enumerate}
  \item[$(i)$] Condition \emph{NA1}$_\mathsf{s}$ holds in the market.
  \item[$(ii)$] $\Def \neq \emptyset$.
  \item[$(iii)$] $S$ is a semimartingale, and $S^i$ cannot revive from bankruptcy for all $i \in \set{1, \ldots, d}$.
  \item[$(iv)$] For all $i \in \set{1, \ldots, d}$, $S^i$ is an exponential semimartingale.
\end{enumerate}
Then, we have the following:
\begin{enumerate}
  \item[(1)] It holds that $(i) \Leftrightarrow (ii) \Rightarrow (iii)$, as well as $(iv) \Rightarrow (i)$.
  \item[(2)] Assume further that $S^i_{\zeta^{S^i} -} > 0$ holds on $\{ \zeta^{S^i} < \infty\}$ for all $i \in \set{1, \ldots, d}$. Then, we have the equivalences $(i) \Leftrightarrow (ii) \Leftrightarrow (iii) \Leftrightarrow (iv)$.
\end{enumerate}

\end{thm}

\subsection{Proof of Theorem \ref{thm: FTAP_jump}, statement $(1)$}

\begin{proof}[$(i) \Rightarrow (ii)$]

Define the set of dyadic rational numbers $\dya := \{ m / 2^k \such k \in \Natural, \, m \in \Natural \}$, which is dense in $\Real_+$. Further, for $k \in \Natural$, define the set of trading times $\parti^k := \{ m / 2^k \such m \in \Natural, \, 0 \leq m \leq k 2^k \}$. Then, $\parti^k \subset \parti^{k'}$ for $k < k'$ and $\bigcup_{k \in \Natural} \parti^k = \dya$. In what follows, $\X^k (1)$ denotes the subset of $\X(1)$ consisting of wealth processes where trading only may happen at times in $\parti^k$. We now state and prove an intermediate result that will help to establish implication $(i) \Rightarrow (ii)$ of Theorem \ref{thm: FTAP_jump}.

\begin{lem} \label{lem: num exists for discrete-time}
Under condition \emph{\naone}, and for each $k \in \Natural$, there exists a wealth process $\tx^k \in \X^k(1)$ with $\prob[\tx^k_t > 0] = 1$ for all $t \in \parti^k$ such that, by defining $\td^k := 1 / \tx^k$, $\expec[\td^k_t X_t  \such \F_s] \leq \td^k_s X_s$ holds for all $X \in \X^k(1)$, where $\parti^k \ni s \leq t \in \parti^k$.
\end{lem}

\begin{proof}
The existence of such ``\num \ portfolio'' $\tx^k$ essentially follows from Theorem 4.12 of \cite{MR2335830}. However, we shall explain in detail below how one can \emph{obtain} the validity of Lemma \ref{lem: num exists for discrete-time} following the idea used to prove Theorem 4.12 of \cite{MR2335830} in this simpler setting, rather than using the latter heavy result. Throughout the proof we keep $k \in \Natural$ fixed, and we set $\parti^k_{++} := \parti^k \setminus \set{0}$.

First of all, it is straightforward to check that condition  \naone \ implies that each $X \in \X$, and in particular also each $S^i$, $\iin$, cannot revive from bankruptcy. This implies that we can consider an alternative ``multiplicative'' characterization of wealth processes in $\X(1)$, as we now describe. Consider a process $\pi = (\pi_t)_{t \in \parti^k_{++}}$ such that,  for all $t \in \parti^k_{++}$, $\pi_t \equiv (\pi^i_t)_{\iin}$ is $\F_{t - 1/2^k}$-measurable and takes values in the $d$-dimensional simplex $\triangle^d := \big\{ z = (z^i)_{i =1, \ldots, d} \in \Real^d \such z^i \geq 0 \text{ for } i=1, \ldots, d, \text{ and } \sum_{i=1}^d z^i \leq 1 \big\}$. Define $X^{(\pi)}_0 := 1$ and, for all $t \in \parti^k_{++}$, $X^{(\pi)}_t := \prod_{\parti^k_{++} \ni u \leq t} \pare{1 + \inner{\pi_u}{\Delta R^{k}_u}}$, where, for $u \in \parti^k_{++}$, $\Delta R^k_u = (\Delta R^{k,i}_u)_{\iin}$ is such that $\Delta R^{k, i}_u = \big( S^i_u / S^i_{u - 1/2^k} - 1 \big) \indic_{\{ S^i_{u - 1/2^k} > 0 \}}$ for $\iin$. Then, define a simple predictable $d$-dimensional process $\theta$ as follows: for $\iin$ and $u \in ]t - 1/2^k, t]$, where $t \in \parti^k_{++}$, set $\theta^i_u = \big( \pi^i_t X^{(\pi)}_{t - 1/2^k} / S^i_{t - 1/2^k} \big) \indic_{\{S^i_{t - 1/2^k}  > 0\}}$; otherwise, set $\theta = 0$. It is then straightforward to check that $X^{1, \theta}$, in the notation of \eqref{eq: Xhat}, is an element of $\X^k(1)$, as well as that $X^{1, \theta}_t = X^{(\pi)}_t$ holds for all $t \in \parti^k$. We have then established that $\pi$ generates a wealth process in $\X^k(1)$. We claim that every wealth process of $\X^k(1)$ can be generated this way. Indeed, starting with any predictable $d$-dimensional process $\theta$ such that $X^{1, \theta}$, in the notation of \eqref{eq: Xhat}, is an element of $\X^k(1)$, we define $\pi^i_t = \big( \theta^i_t S^i_{t - 1/2^k} /X^{1, \theta}_{t - 1/2^k} \big) \indic_{\{X^{1, \theta}_{t - 1/2^k}  > 0\}}$ for $\iin$ and $t \in \parti^k_{++}$. Then, $\pi = (\pi_t)_{t \in \parti^k_{++}}$ is $\triangle^d$-valued, $\pi_t \equiv (\pi^i_t)_{\iin}$ is $\F_{t - 1/2^k}$-measurable for $t \in \parti^k_{++}$, and $\pi$ generates $X^{1, \theta}$ in the way described previously --- in particular, $X^{1, \theta}_t = X^{(\pi)}_t$ holds for all $t \in \parti^k$. (In establishing the claims above it is important that all wealth processes of $\X$ cannot revive from bankruptcy.)

Continuing, since $\triangle^d$ is a \emph{compact} subset of $\Real^d$, for all $t \in \parti^k$ there exists a $\F_{t - 1/2^k}$-measurable $\rho_t = (\rho^i_t)_{\iin}$ such that, for all $\F_{t - 1/2^k}$-measurable and $\triangle^d$-valued $\pi_t = (\pi^i_t)_{\iin}$, we have
\[
\expec \bra{ \frac{ 1 + \inner{\pi_t}{\Delta R^{k}_t}}{1 + \inner{\rho_t}{\Delta R^{k}_t}} \ \Bigg| \ \F_{t - 1/2^k}} \leq 1
\]
(It is exactly the existence of such $\rho_t$ can be seen as a stripped-down version of Theorem 4.12 in \cite{MR2335830}; in effect, $\rho_t$ is the optimal proportions of wealth connected with the log-utility maximization problem, modulo technicalities arising when the value of the log-utility maximization problem has infinite value.) Setting $\tx^k$ to be the wealth process in $\X^k(1)$ generated by $\rho$ as described in the previous paragraph, the result of Lemma \ref{lem: num exists for discrete-time} is immediate.
\end{proof}

We proceed with the proof of implication $(i) \Rightarrow (ii)$ of Theorem \ref{thm: FTAP_jump}, using the notation from the statement of Lemma \ref{lem: num exists for discrete-time}. For all $k \in \Natural$, $\td^k$ satisfies $\td^k_0 = 1$ and is a positive supermartingale when sampled from times in $\parti^k$, since $1 \in \X^k$. Therefore, for any $t \in \dya$, the \emph{convex hull} of the set $\{ \td^k_t  \such k \in \Natural\}$ is bounded in probability. We also claim that, under condition \naone, for any $t \in \Real_+$, the convex hull of the set $\{ \td^k_t  \such k \in \Natural \}$ is bounded away from zero in probability. Indeed, for any collection $(\alpha^k)_{k \in \Natural}$ such that $\alpha^k \geq 0$ for all $k \in \Natural$, having all but a finite number of $\alpha^k$'s non-zero and satisfying $\sum_{k =1 }^\infty \alpha^k = 1$, we have
\[
\frac{1}{\sum_{k = 1}^\infty \alpha^k \td^k } \ \leq \ \sum_{k = 1}^\infty \alpha^k \frac{1}{\td^k} \ = \ \sum_{k = 1}^\infty \alpha^k \tx^k \, \in \, \X (1).
\]
Since, by Proposition \ref{prop: NAone iff NUPBR}, $\set{X_t \such X \in \X (1)}$ is bounded in probability for all $t \in \Real_+$, the previous fact proves that the convex hull of the set $\{ \td^k_t \such k \in \Natural \}$ is bounded away from zero in probability.

Now, using Lemma A1.1 of \cite{MR1304434}, one can proceed as in the proof of Lemma 5.2(a) in \cite{MR1469917} to infer the existence of a sequence $(\widehat{Y}^k)_{k \in \Natural}$ and some process $(\widehat{Y}_t)_{t \in \dya}$ such that, for all $k \in \Natural$, $\widehat{Y}^k$ is a convex combination of $\td^k , \td^{k+1}, \ldots$,  and $\prob[ \lim_{k \to \infty} \widehat{Y}_t^k = Y_t, \, \forall t \in \dya ] = 1$. The discussion of the preceding paragraph ensures that $\prob[0 < \widehat{Y}_t < \infty, \, \forall \, t \in \dya] = 1$.

Let $\dya \ni s \leq t \in \dya$. Then, $s \in \parti^k$ and $t \in \parti^k$ for all large enough $k \in \Natural$. According to the conditional version of Fatou's Lemma, for all $X \in \bigcup_{k=1}^\infty \X^k$ we have that
\begin{equation} \label{eq: Yhat is supermart defl}
\expec [\widehat{Y}_t X_t \such \F_s] \leq \liminf_{k \to \infty} \expec [ \widehat{Y}^k_t X_t \such \F_s] \leq \liminf_{k \to \infty} \widehat{Y}^k_s X_s = \widehat{Y}_s X_s.
\end{equation}
It follows that $(\widehat{Y}_t X_t)_{t \in \dya}$ is a supermartingale  for all $X \in \bigcup_{k=1}^\infty \X^k$. (Observe here that we sample the process $\widehat{Y} X$ only at times contained in $\dya$.) In particular, $(\widehat{Y}_t)_{t \in \dya}$ is a supermartingale.

For any $t \in \Real_+$ define $Y_t := \lim_{s \downarrow \downarrow t, s \in \dya} \widehat{Y}_s$ --- the limit is taken in the $\prob$-a.s. sense, and exists in view of the supermartingale property of $(\widehat{Y}_t)_{t \in \dya}$. It is straightforward that $Y$ is a \cadlag \ process; it is also adapted because $(\F_t)_{t \in \Real_+}$ is right-continuous. Now, for $t \in \Real_+$, let $T \in \dya$ be such that $T > t$; a combination of the right-continuity of both $Y$ and the filtration $(\F_t)_{t \in \Real_+}$, the supermartingale property of  $(\widehat{Y}_t)_{t \in \dya}$, and L\'evy's martingale convergence Theorem, give $\expec[\widehat{Y}_T \such \F_t] \leq Y_t$. Since $\prob[\widehat{Y}_T > 0] = 1$, we obtain $\prob[Y_t > 0] = 1$. Right-continuity of the filtration $(\F_t)_{t \in \Real_+}$, coupled with \eqref{eq: Yhat is supermart defl}, imply that $\expec [Y_t X_t \such \F_s] \leq  Y_s X_s$ for all $\Real_+ \ni s \leq t \in \Real_+$ and $X \in \bigcup_{k=1}^\infty \X^k$. In particular, $Y$ is a \cadlag \ nonnegative supermartingale; since $\prob[Y_t > 0] = 1$ holds for all $t \in \Real_+$, we conclude that $\prob[Y_t > 0, \, \forall t \in \Real_+] = 1$.

Of course, $1 \in \X^k$ and $S^i \in \X^k$ hold for all $k \in \Natural$ and $i \in \set{1, \ldots, d}$. It follows that $Y$ is a supermartingale, as well as that $Y S^i$ is a supermartingale for all $i \in \set{1, \ldots, d}$. In particular, $Y$ and $Y S = (Y S^i)_{i \in \set{1, \ldots, d}}$ are semimartingales. Consider any $X^{x, \theta}$ in the notation of \eqref{eq: Xhat}. Using the integration-by-parts formula, we obtain
\[
 Y X^{x, \theta} = x + \int_0^\cdot \pare{ X^{x, \theta}_{t-} - \inner{\theta_t}{S_{t-}}} \ud Y_t + \int_0^\cdot \inner{\theta_t}{\ud (Y_t S_t)}.
\]
If $X^{x, \theta} \in \X(x)$, we have $X^{x, \theta}_{-} - \inner{\theta}{S_-} \geq 0$, as well as $\theta^i \geq 0$ for $i \in \set{1, \ldots, d}$. Then, the supermartingale property of $Y$ and $Y S^i$, $i \in \set{1, \ldots, d}$, gives that $Y X^{x, \theta}$ is a supermartingale. Therefore, $Y \in \Def$, i.e., $\Def \neq \emptyset$.
\end{proof}

\begin{proof}[$(ii) \Rightarrow (i)$]
Let $Y \in \Def$, and fix $T \in \Real_+$. Then, $\sup_{X \in \X (1)} \expec[Y_T X_T] \leq 1$. In particular, the set $\set{Y_T X_T \such X \in \X (1)}$ is bounded in probability. Since $\prob[Y_T > 0] = 1$, the set $\set{X_T \such X \in \X (1)}$ is bounded in probability as well. An invocation of Proposition \ref{prop: NAone iff NUPBR} finishes the argument.
\end{proof}

\begin{proof}[$(ii) \Rightarrow (iii)$]
Let $Y \in \Def$. Since $S^i \in \X$, $Y S^i$ is a supermartingale, thus a semimartingale, for all $i \in \set{1, \ldots, d}$. Also, the fact that $Y > 0$ and It\^o's formula give that $1 / Y$ is a semimartingale. Therefore, $S^i = (1 / Y) (Y S^i)$ is a semimartingale for all $i \in \set{1, \ldots, d}$. Furthermore, since $Y S^i$ is a nonnegative supermartingale, we have $Y_t S^i_t = 0$ for all $t \geq \zeta^{S^i}$ on $\{\zeta^{S^i} < \infty \}$, for $i \in \set{1, \ldots, d}$. Now, using $Y > 0$ again, we obtain that $S^i_t = 0$ holds for all $t \geq \zeta^{S^i}$ on $\{ \zeta^{S^i} < \infty \}$. In other words, each $S^i$, $i \in \set{1, \ldots, d}$, cannot revive after bankruptcy.
\end{proof}

\begin{proof}[$(iv) \Rightarrow (i)$]
Since $S$ is a semimartingale, we can consider continuous-time trading. For $x \in \Real_+$, let $\oX (x)$ be the set of all  wealth processes $X^{x, \theta} := x + \int_0^\cdot \inner{\theta_t}{\ud S_t}$, where $\theta$ is $d$-dimensional, predictable and $S$-integrable, ``$\int_0^\cdot \inner{\theta_t}{\ud S_t}$'' denotes a vector stochastic integral, $X^{x, \theta} \geq 0$ and $0 \leq \inner{\theta}{S_-} \leq X_-^{x, \theta}$. (Observe that the qualifying subscript ``$\mathsf{s}$'' denoting simple trading has been dropped in the definition of $\X(x)$, since we are considering continuous-time trading.) Of course, $\X(x) \subseteq \oX(x)$. We shall show in the next paragraph that $\set{X_T \such X \in \oX(1)}$ is bounded in probability for all $T \in \Real_+$, therefore establishing condition \naone, in view of Proposition \ref{prop: NAone iff NUPBR}.

For all $i \in \set{1, \ldots, d}$, write $S^i = S^i_0 \Exp(R^i)$, where $R^i$ is a semimartingale with $R^i_0 = 0$. Let $R := (R^i)_{i=1, \ldots, d}$. It is straightforward to see that $\oX (1)$ coincides with the class of all processes of the form $\Exp \pare{ \int_0^\cdot \inner{\pi_t}{\ud R_t} }$, where $\pi$ is predictable and take values in the $d$-dimensional simplex $\triangle^d := \big\{ z = (z^i)_{i =1, \ldots, d} \in \Real^d \such z^i \geq 0 \text{ for } i=1, \ldots, d, \text{ and } \sum_{i=1}^d z^i \leq 1 \big\}$. Since, for all $T \in \Real_+$,
\[
\log \pare{\Exp \pare{ \int_0^T \inner{\pi_t}{\ud R_t} }} \leq \int_0^T \inner{\pi_t}{\ud R_t}
\]
holds for all $\triangle^d$-valued and predictable $\pi$, it suffices to show the boundedness in probability of the class of all $\int_0^T \inner{\pi_t}{\ud R_t}$, where $\pi$ ranges in all $\triangle^d$-valued and predictable processes. Write $R = B + M$, where $B$ is a process of finite variation and $M$ is a local martingale with $|\Delta M^i| \leq 1$, $i \in \set{1, \ldots, d}$. Then, $\int_0^T |\inner{\pi_t}{\ud B_t}| \leq \sum_{i =1}^d \int_0^T |\ud B^i_t| < \infty$. This establishes the boundedness in probability of the class of all $\int_0^T \inner{\pi_t}{\ud B_t}$, where $\pi$ ranges in all $\triangle^d$-valued and predictable processes. We have to show that the same holds for the class of all $\int_0^T \inner{\pi_t}{\ud M_t}$, where $\pi$ is $\triangle^d$-valued and predictable. For $k \in \Natural$, let $\tau^k := \inf \{ t \in \Real_+ \such \sum_{i =1}^d [M^i, M^i]_t \geq k \} \wedge T$, Note that $[ M^i, M^i]_{\tau^k} = [ M^i, M^i]_{\tau^k-} + |\Delta M^i_{\tau^k}|^2 \leq  k + 1$ holds for all $i \in \set{1, \ldots, d}$. Therefore, using the notation $\normtwo{\eta} := \sqrt{\expec[|\eta|^2]}$ for a random variable $\eta$, we obtain
\[
 \normtwo{\int_0^{\tau^k} \inner{\pi_t}{\ud M_t}} \leq \sum_{i=1}^d \normtwo{ \int_0^{\tau^k} \pi^i_t \ud M^i_t  } \leq \sum_{i=1}^d \normtwo{  \sqrt{[ M^i, M^i]_{\tau^k}}} \leq d \sqrt{k + 1}
\]
Fix $\epsilon > 0$. Let $k = k (\epsilon)$ be such that $\prob[\tau^k < T] < \epsilon / 2$, and also let $\ell := d \sqrt{2 (k + 1) / \epsilon} $. Then,
\[
 \prob \bra{ \int_0^T \inner{\pi_t}{\ud M_t} > \ell} \leq \prob \bra{\tau^k < T} + \prob \bra{ \int_0^{\tau^k} \inner{\pi_t}{\ud M_t} > \ell} \leq \frac{\epsilon}{2} + \abs{\frac{\normtwo{ \int_0^{\tau^k} \inner{\pi_t}{\ud M_t}}}{\ell}}^2 \leq \epsilon.
\]
The last estimate is uniform over all $\simplex$-valued and predictable $\pi$. We have, therefore, established the boundedness in probability of the class of all $\int_0^T \inner{\pi_t}{\ud M_t}$, where $\pi$ ranges in all $\triangle^d$-valued and predictable processes. This completes the proof.
\end{proof}

\subsection{Proof of Theorem \ref{thm: FTAP_jump}, statement $(2)$.}

In view of statement (1) of Theorem \ref{thm: FTAP_jump}, we only need to show the validity of $(iii) \Leftrightarrow (iv)$ under the extra assumption of statement (2). This equivalence is really Proposition 2.2 in \cite{MR2322919}, but we present the few details for completeness.

For the implication $(iii) \Rightarrow (iv)$, simply define $R^i := \int_0^\cdot (1 / S^i_{t-}) \ud S^i_t$ for $i \in \set{1, \ldots, d}$,  The latter process is a well-defined semimartingale because, for each $\iin$, $S^i$ is a semimartingale, $S^i_-$ is locally bounded away from zero on the stochastic interval $\dbra{0, \zeta^{S^i}}$, and $S = 0$ on $\dbraco{\zeta^{S^i}, \infty}$.

Now, for $(iv) \Rightarrow (iii)$, it is clear that $S$ is a semimartingale. Furthermore, for all $i \in \set{1, \ldots, d}$, $S^i$ cannot revive from bankruptcy; this follows because stochastic exponentials stay at zero once they hit zero.
\qed

\section{On and Beyond the Main Result} \label{sec: beyond main result}

\subsection{Comparison with the result of Delbaen and Schachermayer} \label{subsubsec: compar with DS}

Theorem 7.2  of the seminal paper \cite{MR1304434} establishes the semimartingale property of $S$ under condition NFLVR for simple admissible strategies, coupled with a local boundedness assumption on $S$ (always together with the \cadlag \ property and adaptedness). The assumptions of Theorem \ref{thm: FTAP_jump} are different than the ones in \cite{MR1304434}. Condition \naone \ (valid for simple, no-short-sales trading) is weaker than NFLVR for simple admissible strategies. Furthermore, local boundedness from above is not required in our context, but we do require that each $S^i$, $i \in \set{1, \ldots, d}$, is nonnegative. In fact, as we shall argue in \S \ref{subsubsec: local bdd from below} below, nonnegativity of each $S^i$, $i \in \set{1, \ldots, d}$, can be weakened by local boundedness from below, indeed making Theorem \ref{thm: FTAP_jump} a generalization of Theorem 7.2 in \cite{MR1304434}. Note that if the components of $S$ are unbounded both above and below, not even condition NFLVR is enough to ensure the semimartingale property of $S$; see Example 7.5 in \cite{MR1304434}.

\smallskip
Interestingly, and in contrast to \cite{MR1304434}, the proof of Theorem \ref{thm: FTAP_jump} provided here does \emph{not} use the deep Bichteler-Dellacherie theorem on the characterization of semimartingales as ``good integrators'' (see \cite{MR1906715}, \cite{MR2273672}, where one \emph{starts} by defining semimartingales as good integrators and obtains the classical definition as a by-product). Actually, and in view of Proposition \ref{prop: NAone iff NUPBR}, statement (2) of Theorem \ref{thm: FTAP_jump} can be seen as a ``multiplicative'' counterpart of the Bichteler-Dellacherie theorem. Its proof exploits two simple facts: (a) positive supermartingales are semimartingales, which follows directly from the Doob-Meyer decomposition theorem; and (b) reciprocals of strictly positive supermartingales are semimartingales, which is a consequence of It\^o's formula. Crucial in the proof is also the concept of the \num \ portfolio.

\subsection{The semimartingale property of $S$ when each $S^i$, $\iin$, is locally bounded from below} \label{subsubsec: local bdd from below}

As mentioned previously, implication $(i) \Rightarrow (iii)$ actually holds even when each $S^i$, $\iin$,  is locally bounded from below, which we shall establish now. We still, of course, assume that each $S^i$, $\iin$, is adapted and \cadlag. Since ``no-short-sales'' strategies have ambiguous meaning when asset prices can become negative, we need to make some changes in the class of admissible wealth processes. For $x \in \Real_+$, let $\X'(x)$ denote the class of all wealth processes $X^{x, \theta}$ using simple trading as in \eqref{eq: Xhat} that satisfy $X^{x, \theta} \geq 0$. Further, set $\X' = \bigcup_{x \in \Real_+} \X'(x)$. Define condition \naonep \ for the class $\X'$ in the obvious manner, replacing ``$\X$'' with ``$\X'$'' throughout in \S \ref{subsec: UPBR}. Assume then that condition \naonep \ holds. To show that $S$ is a semimartingale, it is enough to show that $(S_{\tau^k \wedge t})_{t \in \Real_+}$ is a semimartingale for each $k \in \Natural$, where $(\tau^k)_{k \in \Natural}$ is a localizing sequence such that $S^i \geq -k$ on $\dbra{0, \tau^k}$ for all $\iin$ and $k \in \Natural$. In other words, we might as well assume that $S^i \geq - k$ for all $i \in \set{1, \ldots, d}$. Define $\tilde{S}^i := k + S^i$; then, $\tilde{S}^i$ is nonnegative for all $\iin$. Let $\tilde{S} = (\tilde{S}^i)_{i \in \set{1, \ldots, d}}$. If $\tilde{\X}$ is (in self-explanatory notation) the collection of all wealth processes resulting from simple, no-short-sales strategies investing in $\tilde{S}$, it is straightforward that $\tilde{\X} \subseteq \X'$. Therefore, \naone \ holds for simple, no-short-sales strategies investing in $\tilde{S}$; using implication $(i) \Rightarrow (iii)$ in statement (1) of Theorem \ref{thm: FTAP_jump}, we obtain the semimartingale property of $\tilde{S}$. The latter is of course equivalent to $S$ being a semimartingale.

\smallskip

One might wonder why we do not simply ask from the outset that each $S^i$, $\iin$, is locally bounded from below, since it certainly contains the case where each $S^i$, $\iin$, is nonnegative. The reason is that by restricting trading to using only no-short-sales strategies (which we can do when each $S^i$, $\iin$, is nonnegative) enables us to be as general as possible in extracting the semimartingale property of $S$ from the \naone \ condition. Consider, for example, the discounted asset-price process given by $S = a \indic_{\dbraco{0,1}} + b \indic_{\dbraco{1,\infty}}$, where $a > 0$ and $b \in \Real_+$ with $a \neq b$. This is a \emph{really} elementary example of a nonnegative semimartingale. Now, if we allow for any form of simple trading, as long as it keeps the wealth processes nonnegative, it is clear that condition \naonep \ will fail (since it is known that at time $t = 1$ there will be a jump of size $(b - a) \in \Real \setminus \set{0}$ in the discounted asset-price process). On the other hand, if we only allow for no-short-sales strategies, \naone \ will hold --- this is easy to see directly using Proposition \ref{prop: NAone iff NUPBR}, since $X_T \leq |b - a| / a$ for all $T \geq 1$ and $X \in \X(1)$. Therefore, we can conclude that $S$ is a semimartingale using implication $(i) \Rightarrow (iii)$ in statement (1) of Theorem \ref{thm: FTAP_jump}. (Of course, one might argue that there is no need to invoke Theorem \ref{thm: FTAP_jump} for the simple example here. The point is that allowing for all nonnegative wealth processes results in a rather weak sufficient criterion for the semimartingale property of $S$.)

\subsection{The semimartingale property of $S$ via bounded indirect utility}
There has been previous work in the literature obtaining the semimartingale property of $S$ using the finiteness of the value function of a utility maximization problem via use of only simple strategies --- see, for instance, \cite{MR2139030}, \cite{MR2157199}, \cite{LarZit05}. In all cases, there has been an assumption of local boundedness (or even continuity) on $S$. We shall offer a result in the same spirit, dropping the local boundedness requirement. We shall assume \emph{either} that discounted asset-price processes are nonnegative and \emph{only} no-short-sales simple strategies are considered (which allows for a sharp result), \emph{or} that discounted asset-price processes are locally bounded from below. In the latter case, Proposition \ref{prop: semimart via finite expec util} that follows is a direct generalization of the corresponding result in \cite{MR2139030}, where the authors consider locally bounded (both above and below) discounted asset-price processes. In the statement of Proposition \ref{prop: semimart via finite expec util} below, we use the notation $\X'(x)$ introduced previously in \S \ref{subsubsec: local bdd from below}.

\begin{prop} \label{prop: semimart via finite expec util}
Let $S = (S^i)_{i =1, \ldots, d}$ be such that $S^i$ is adapted and \cadlag \ process for $i \in \set{1, \ldots, d}$. Also, let $U: \Real_+ \mapsto \Real \cup \{ - \infty \}$ be a nondecreasing function with $U > - \infty$ on $]0 , \infty]$ and $U(\infty) = \infty$. Fix some $x > 0$. Finally, let $T$ be a finite stopping time. Assume that either:
\begin{itemize}
	\item each $S^i$, $\iin$, is nonnegative and $\sup_{X \in \X(x)} \expec[U(X_T)] < \infty$, or 
	\item each $S^i$, $\iin$, is locally bounded from below and $\sup_{X \in \X'(x)} \expec[U(X_T)] < \infty$.
\end{itemize}
Then, the process $(S_{T \wedge t})_{t \in \Real_+}$ is a semimartingale.
\end{prop}

\begin{proof}
Start by assuming that each $S^i$, $\iin$ is nonnegative and that $\sup_{X \in \X(x)} \expec[U(X_T)] < \infty$. Since we only care about the semimartingale property of $(S_{T \wedge t})_{t \in \Real_+}$, assume without loss of generality that $S_t = S_{T \wedge t}$ for all $t \in \Real_+$. Suppose that condition \naone \ fails. According to Proposition \ref{prop: NAone iff NUPBR} and Remark \ref{rem: NAone for stop times}, there exists a sequence $(\tx^n)_{n \in \Natural}$ of elements in $\X(x)$ and $p > 0$ such that $\prob [\tx_T^n > 2 n] \geq p$ for all $n \in \Natural$. For all $n \in \Natural$, let $X^n := (x + \tx^n)/2 \in \X(x)$. Then, $\sup_{X \in \X(x)} \expec[U(X_T)] \geq \liminf_{n \to \infty} \expec[U(X_T^n)] \geq (1 - p) U(x/2) + p \liminf_{n \to \infty} U(n) = \infty$. This is a contradiction to $\sup_{X \in \X(x)} \expec[U(X_T)] < \infty$. We conclude that $(S_{T \wedge t})_{t \in \Real_+}$ is a semimartingale using implication $(i) \Rightarrow (iii)$ in statement (1) of Theorem \ref{thm: FTAP_jump}. 

Under the assumption that each $S^i$, $\iin$ is locally bounded from below and that $\sup_{X \in \X'(x)} \expec[U(X_T)] < \infty$, the proof is exactly the same as the one in the preceding paragraph, provided that one replaces ``$\X$'' with ``$\X'$'' throughout, and uses the fact that condition \naonep \ for the class $\X'$ implies the semimartingale property for $S$, as was discussed in \S \ref{subsubsec: local bdd from below}.
\end{proof}

\subsection{On the implication $(iii) \Rightarrow (i)$ in Theorem \ref{thm: FTAP_jump}}
If we do not require the additional assumption on $S$ in statement (2) of Theorem \ref{thm: FTAP_jump}, implication $(iii) \Rightarrow (i)$ might fail. We present below a counterexample where this happens.

On $(\Omega, \F, \prob)$, let $W$ be a standard, one-dimensional Brownian motion (with respect to its own natural filtration --- we have not defined $(\F_t)_{t \in \Real_+}$ yet). Define the process $\xi$ via $\xi_t := \exp(- t / 4 + W_t)$ for $t \in \Real_+$. Since $\lim_{t \to \infty} W_t / t = 0$, $\prob$-a.s., it is straightforward to check that $\xi_\infty := \lim_{t \to \infty} \xi_t = 0$, and actually that $\int_0^\infty \xi_t \ud t < \infty$, both holding $\prob$-a.s. Write $\xi = A + M$ for the Doob-Meyer decomposition of the continuous submartingale $\xi$ under its natural filtration, where $A = (1 / 4) \int_0^\cdot \xi_t \ud t$ and $M = \int_0^\cdot \xi_t \ud W_t$. Due to $\int_0^\infty \xi_t \ud t < \infty$, we have $A_\infty < \infty$ and $[M, M]_\infty = \int_0^\infty |\xi_t|^2 \ud t < \infty$, where $[M, M]$ is the quadratic variation process of $M$. In the terminology of \cite{MR2126973}, $\xi$ is a semimartingale up to infinity. If we define $S$ via $S_t = \xi_{t / (1 - t)}$ for $t \in [0, 1[$ and $S_t = 0$ for $t \in [1, \infty[$, then $S$ is a nonnegative semimartingale. Define $(\F_t)_{t \in \Real_+}$ to be the augmentation of the natural filtration of $S$. Observe that $\zeta^S = 1$ and $S_{\zeta^S -} = 0$; the condition of statement (2) of Theorem \ref{thm: FTAP_jump} is not satisfied. In order to establish that \naone \ fails, and in view of Proposition \ref{prop: NAone iff NUPBR}, it is sufficient to show that $\set{X_1 \such X \in \X(1)}$ is not bounded in probability. Using continuous-time trading, define a wealth process $\widehat{X}$ for $t \in [0, 1[$, via $\widehat{X}_0 = 1$ and the dynamics $\ud \widehat{X}_t / \widehat{X}_t = (1 / 4) (\ud S_t / S_t)$ for $t \in [0,1[$. Then, $\widehat{X}_t = \exp \pare{ (1/16) (t / (1-t)) + (1/4) W_{t / (1-t)}}$ for $t \in [0,1[$,
which implies that $\prob[\lim_{t \uparrow \uparrow 1} \widehat{X}_t = \infty] = 1$, where ``$t \uparrow \uparrow 1$'' means that $t$ \emph{strictly} increases to $1$. Here, the percentage of investment is $1/4 \in [0,1]$, i.e, $\widehat{X}$ is the result of a no-short-sales strategy. One can then find an approximating sequence $(X^k)_{k \in \Natural}$ such that $X^k \in \X(1)$ for all $k \in \Natural$, as well as $\prob[| X^k_{1}  - \widehat{X}_{1 - 1 / k} | < 1] > 1 - 1 / k$. (Approximation results of this sort are discussed in greater generality in \cite{MR1971602}.) Then, $(X^k_1)_{k \in \Natural}$ is not bounded in probability; therefore,  \naone \ fails. Of course, in this example we also have $(iii) \Rightarrow (iv)$ of Theorem \ref{thm: FTAP_jump} failing.

% ----------------------------------------------------------------
\bibliography{semimarts_in_fin_mod_new}
\bibliographystyle{siam}
\end{document}